\documentclass[journal=ancac3,manuscript=letter]{achemso}
\usepackage{multicol}

\usepackage[T1]{fontenc}
\usepackage[utf8]{inputenc}
\usepackage{babel}
\usepackage{amsmath}
\usepackage{amsthm}
\usepackage{amssymb}
\usepackage{graphicx}
\usepackage{xargs}[2008/03/08]
\usepackage[unicode=true, bookmarks=false,
 breaklinks=false,pdfborder={0 0 1},backref=false,colorlinks=false]
 {hyperref}
\hypersetup{pdftitle=pdfauthor={R. Gutierrez} }
 
 \newcommand{\change}[1]{{\color{black} #1}}
 
\makeatletter

\title[\textsf{hel}]{The Role of Exchange Interactions in the Magnetic Response and Inter-Molecular Recognition of Chiral Molecules}

\usepackage{braket}
\usepackage{tikz-cd}

\author{Arezoo Dianat}
\email{arezoo.dianat@tu-dresden.de}
 \affiliation[]{Institute for Materials Science and Max Bergmann Center of Biomaterials, TU Dresden, 01062 Dresden, Germany}
\author{Rafael Gutierrez}
\email{rafael.gutierrez@tu-dresden.de}
\affiliation[]{Institute for Materials Science and Max Bergmann Center of Biomaterials, TU Dresden, 01062 Dresden, Germany}

\author{Hen Alpern}
\affiliation[]{Applied Physics Department and the Center for Nanoscience and Nanotechnology, The Hebrew University of Jerusalem, Jerusalem 9190401, Israel}
\alsoaffiliation[]{Racah Institute of Physics and the Center for Nanoscience and Nanotechnology, The Hebrew University of Jerusalem, Jerusalem 9190401, Israel}

\author{Vladimiro Mujica}
\affiliation[]{School of Molecular Sciences, Arizona State University, Tempe, Arizona, 85287, USA}
\alsoaffiliation[]{Ikerbasque Foundation and Donostia International Physics Center (DIPC), Manuel de Lardizabal Pasealekua 4, 20018 Donostia, Euskadi, Spain}

\author{Amir Ziv}
\affiliation[]{Applied Physics Department and the Center for Nanoscience and Nanotechnology, The Hebrew University of Jerusalem, Jerusalem 9190401, Israel}

\author{Shira Yochelis}
\affiliation[]{Applied Physics Department and the Center for Nanoscience and Nanotechnology, The Hebrew University of Jerusalem, Jerusalem 9190401, Israel}

\author{Oded Millo}
\affiliation[]{Racah Institute of Physics and the Center for Nanoscience and Nanotechnology, The Hebrew University of Jerusalem, Jerusalem 9190401, Israel}

\author{Yossi Paltiel}
\affiliation[]{Applied Physics Department and the Center for Nanoscience and Nanotechnology, The Hebrew University of Jerusalem, Jerusalem 9190401, Israel }

\author{Gianaurelio Cuniberti}
\email{gianaurelio.cuniberti@tu-dresden.de}
\affiliation[]{Institute for Materials Science and Max Bergmann Center of Biomaterials, TU Dresden, 01062 Dresden, Germany}
\alsoaffiliation[]{Dresden Center for Computational Materials Science (DCMS), TU Dresden, 01062 Dresden, Germany}
\SectionNumbersOn

\makeatother

\usepackage{babel}
\begin{document}

\maketitle

%\newpage

\begin{abstract}
The physical origin  of so called Chirality-Induced Spin Selectivity (CISS) effect has puzzled experimental and theoretical researchers over the past few years. Early experiments were interpreted in terms of  unconventional spin-orbit interactions mediated by the helical geometry. However, more recent experimental studies have clearly revealed that electronic exchange interactions also play a  key role in the magnetic response  of chiral molecules \change{in singlet states}. In this  investigation, we use spin polarized closed-shell Density-Functional Theory calculations to address the influence of exchange contributions to the interaction between helical molecules as well as of helical molecules with magnetized  substrates. We show that exchange effects result in differences in the interaction properties with magnetized surfaces, shedding light into the possible origin of two recent  important experimental results: enantiomer separation  and  Magnetic Exchange Force Microscopy with AFM tips functionalized with helical peptides. \\


\end{abstract}

%%%%%%%%%%%%%%%%%%%%%%%%%%%%%%%%%%%%%%%%%%%%%%%%%%%%%%%%%%%%%%%%%%%
%\textbf{Keywords}: CISS effect, Density-Functional Theory, Exchange Effects, Helical Molecules \\

%\begin{multicols}{2}

%\section*{Introduction}
The close, though rather unexpected, interconnection between  chiral symmetry and spin-dependent transfer processes in molecular systems -an effect denoted meanwhile as Chirality-Induced Spin Selectivity (CISS)-  has been puzzling experimental and theoretical researchers over at least a decade. \change{The} first experimental results for self-assembled  monolayers suggesting  this spin-chirality connection were   published in 1999;~\cite{Ray99} however, it is probably in the work of J. Kessler and others,~\cite{kessler} where first hints can be found on what was called  electron dichroism, i.e. a spin sensitivity of the electron scattering cross section of chiral molecules. However, the measured effects were far too small at the time. In 2011, photoemission experiments, gaining direct access to the electron spin polarization, demonstrated a very large spin selectivity  in DNA molecules, with spin polarizations reaching values as high as 60 \% and with a linear dependence of the polarization with  molecular length.~\cite{Gohler2011} This strong spin selectivity   was shortly hereafter corroborated in two-point electrical transport measurements in DNA.~\cite{doi:10.1021/nl2021637} 
Following these experimental results, the CISS effect has been broadly demonstrated in large classes of molecular systems, including Bacteriorhodopsin,~\cite{Mishra14872} various oligopeptides,~\cite{doi:10.1063/1.4966237,doi:10.1021/acsnano.5b00832,SMLL:SMLL201602519,doi:10.1021/jp509974z,doi:10.1021/jacs.8b08421,doi:10.1021/acs.jpclett.9b03487,C9CP04681J} DNA,~\cite{doi:10.1021/jacs.6b10538} DNA-wrapped carbon nanotubes,~\cite{C6NR09395G} helicenes,~\cite{ADMA:ADMA201504725,doi:10.1021/acs.jpclett.8b00208} and supramolecular nanofibers.~\cite{doi:10.1002/adma.201904965} Further, additional CISS-related effects have been shown, including magnetization switching induced by helical molecular films~\cite{BenDor2017} as well as its relevance for the electrochemical  water splitting process,~\cite{doi:10.1021/jacs.6b12971} and for short-range intermolecular interactions in biosystems.~\cite{Kumar2017} For an overview of the field we refer the reader to various review articles.~\cite{C6CS00369A,doi:10.1146/annurev-physchem-040214-121554}   

From the theoretical perspective, many studies have been carried out using model Hamiltonian approaches in an attempt to rationalize the CISS effect.~\cite{Yeganeh09,Medina12,Gutierrez12,Gutierrez13,PhysRevB.88.165409,Medina15,Guo12,Guo14a,Guo14b,doi:10.1021/acs.jpcc.9b05020,Matityahu16,Caetano16,Diaz17,Diaz17bis,PhysRevB.95.085411,doi:10.1063/1.4820907,doi:10.1021/acs.jpclett.8b02196,doi:10.1021/acs.nanolett.9b01707,Geyer2019,doi:10.1021/acs.jpclett.9b02929,Varela17,shitade2020,paltiel2020,geyer2020,utsumi20,biom10010049} 
Independently of the specific details, the vast majority of these models establish a connection between the spin selectivity and the presence of  spin-orbit interaction related to the helical molecular shape (a different approach has been taken, however, in Refs.~\cite{doi:10.1021/acs.nanolett.9b01707,doi:10.1063/1.4820907}, stressing more the influence of the molecule-metal interface). The computed spin polarizations are in general rather small, unless some type of symmetry breaking is introduced, related to e.g. B\"uttiker probes,~\cite{Guo12,Guo14a,Guo14b} leakages,~\cite{Matityahu16} or time-reversal symmetry.\cite{doi:10.1021/acs.jpcc.9b05020,geyer2020} Atomistic first-principle based investigations  are, on the contrary, still rare.~\cite{doi:10.1021/acs.jpclett.8b02360,doi:10.1021/acs.jpclett.8b00208,carmen} Thus, despite considerable progress in elucidating the origin of the CISS,  the problem is still subject to intensive debate. 

More recently, several experimental studies have started to highlight the influence of electronic exchange effects in the spin-dependent magnetic response of helical molecules and in molecular recognition events. These exchange effects are part of the molecular magnetic response and under some conditions might be dominant over spin-orbit coupling, \change{an effect that can be understood given the complex dependence of a many-electron state of a molecular system both on the spin-orbit coupling and the  spin symmetry requirement on the state imposed by Pauli Principle. In fact, even if spin-orbit interaction is neglected, spin enter into the description of molecular interactions via the antisymmetry requirement on the wavefunction, which leads to the appearance of exchange interactions}. In a ground-breaking experiment, enantiomer separation mediated by exchange interactions was demonstrated by Banerjee-Ghosh et al.~\cite{Banerjee-Ghosh1331}   Selective surface adsorption of chiral molecules was observed as a result of the difference in the kinetic adsorption rates associated with the enantiomers, the effect becoming much weaker at longer times due to the dominance of thermodynamic effects. These results were rationalized based on the assumption of a transient chirality-dependent spin polarization in the molecules while approaching the substrate, leading to a    chirality-dependent  molecule-substrate exchange interaction (see also Ref.~\cite{doi:10.1063/1.5125034} for a similar effect in a different chiral peptide and Ref.~\cite{doi:10.1021/acs.jpcb.9b07987} concerning the interplay between substrate magnetization and an applied electric field on the molecular adsorption rates). The results presented by Banerjee-Ghosh et al. pointed for the first time, to the best of our knowledge, at the  influence of exchange-mediated processes, which had been neglected so far in the description of the magnetic response of helical molecular systems. In another study, A. Ziv et al.~\cite{doi:10.1002/adma.201904206} introduced a novel version of a Magnetic Exchange Force Microscope by functionalizing the AFM tip with a helical peptide. 
Using this approach, a dependence of  the pulling force on the magnetization direction of a Nickel substrate was demonstrated. This effect was especially clear in the obtained force histograms, which showed two prominent peaks in the distribution, one peak common to both directions of the magnetic moment of the substrate as well as of the non-magnetic substrate, and another one whose width did  depend on the substrate magnetic state and was absent with no magnetic field applied to the substrate. 
Theoretically,  Ref.~\cite{doi:10.1021/acs.jpclett.9b02929} is the first  study combining Hubbard correlations and spin-orbit in a model Hamiltonian approach to spin transport in helical molecules and highlighting the importance of both types of interactions. In fact, it is becoming clear that the magnetic response of chiral molecules must include the CISS and the exchange contributions, both of them being spin dependent. 

Motivated by the results in Refs.~\cite{Banerjee-Ghosh1331,doi:10.1063/1.5125034,doi:10.1002/adma.201904206}, we address in this computational study based on Density-Functional theory (DFT), the influence of exchange interactions in the magnetic response and molecular recognition of chiral molecules by magnetic substrates. For this, we will study arrays of helical molecules interacting with  magnetized substrates. Obviously, the problem at stake is extremely complex, so that we are going to consider rather idealized situations as described in the following sections.  

We show that below a certain critical intermolecular distance in an \change{model} molecular  assembly far away from  a substrate, an insulator to metal transition takes place, accompanied by the emergence of a non-zero spin density around the Fermi energy. This spin symmetry breaking effect bears some similarity with the so called singlet magnetic ground state, where a broken spin symmetry (BSS) state emerges as lowest energy state while still remaining a singlet. \change{This BSS state, where global singlet symmetry is preserved, while locally the spin densities are non zero, has been described e.g. in gold nanoparticles to which sulfur linkers were adsorbed.~\cite{doi:10.1021/jp054583g,doi:10.1021/jz201326k} Due to the absence of any external spin-dependent perturbations in the ideal molecular array, time-reversal symmetry is preserved and energetically degenerate solutions with positive as well as negative net spin density around the Fermi energy are possible for each (L- or D-) enantiomer. }

By approaching a magnetized substrate, \change{the magnitude of the net magnetization around the Fermi level is reduced, but is still relevant in the range of weak to strong physisorption}. Upon covalent bonding to the substrate, in the chemisorption regime, the metallic-like state is \change{weakened}, a pseudo-gap is visible,  but still a small,  non-zero spin density remains around the Fermi energy.
In the second part of this study, we focus on the setup discussed in Ref.~\cite{doi:10.1002/adma.201904206} and show that the interaction of a helical molecule with an up or down magnetized substrate sensitively depends on the substrate magnetization direction, in good agreement with the experimental results. We trace back this result to  exchange-dependent structural relaxation effects, \change{which are now able to break the previously mentioned degeneracy. Overall, we  believe that our study contributes to shed light on the possible role of exchange effects in the description of intermolecular interactions in chiral molecules and on their interaction with magnetized surfaces.}

%\section{Results}
 The calculations were carried out using spin-polarized density functional theory (DFT) with the Perdew–Burke-Ernzerhof (PBE) generalized gradient approximation (GGA) for
the exchange-correlation functional, and the augmented-plane-wave (PAW) method using the Vienna \textit{ab initio} simulation package (VASP). Some of our results were further checked using the  strongly constrained and
appropriately normed (SCAN) meta-GGA functional,~\cite{scan} \change{as well as with a GGA+U functional (with $U$=5 eV and $J$=0.1 eV).~\cite{PhysRevB.62.16392} In all cases, the results  show the closing of the band gap and the onset of a broken spin symmetry (BSS)  state, as discussed below, demonstrating the robustness of our results.} Therefore, we will only discuss, \change{in what follows}, the PBE-based results. 
Wave functions were expanded in plane waves up to a kinetic energy cutoff of 400 eV. \change{Additional checks were performed using a cutoff of 800 eV in the case of L-polyglycine and, although the total energy changed by around 0.2 eV, the difference in energy between the broken spin symmetry state and the unbroken symmetry situation remained the same. }
Integration in the first Brillouin zone was performed using the Monkhorst-Pack grids, including 90 k-points in the primitive cell. For all considered structures, the  atomic coordinates were fully optimized until all force components were less than 0.01 eV/\AA. Convergence of energy differences with respect to the used cutoff energies and $k$-point grids were tested in all cases within a tolerance of 
10 meV/atom. The periodically repeated simulation cells included slabs with three substrate layers of a 4x4 supercell of Ni(111).  In all cases, the vacuum gap between the slab surface models is larger than 15 \AA. Dispersion corrections were included through the standard D2 Grimme parametrization.~\cite{grimme}
We remark that no spin-orbit interaction was considered in the presented calculations, so that all the spin dependence of our results arises because of the antisymmetry requirements in the total electronic wave function. The coexistence of spin-orbit interactions and exchange effects remains open to future investigations, \change{and is expected to play an important role for the description of generalized van der Waals interactions between chiral molecules}.

\begin{figure}[t!]
    \centering
    \includegraphics[width=0.8\textwidth]{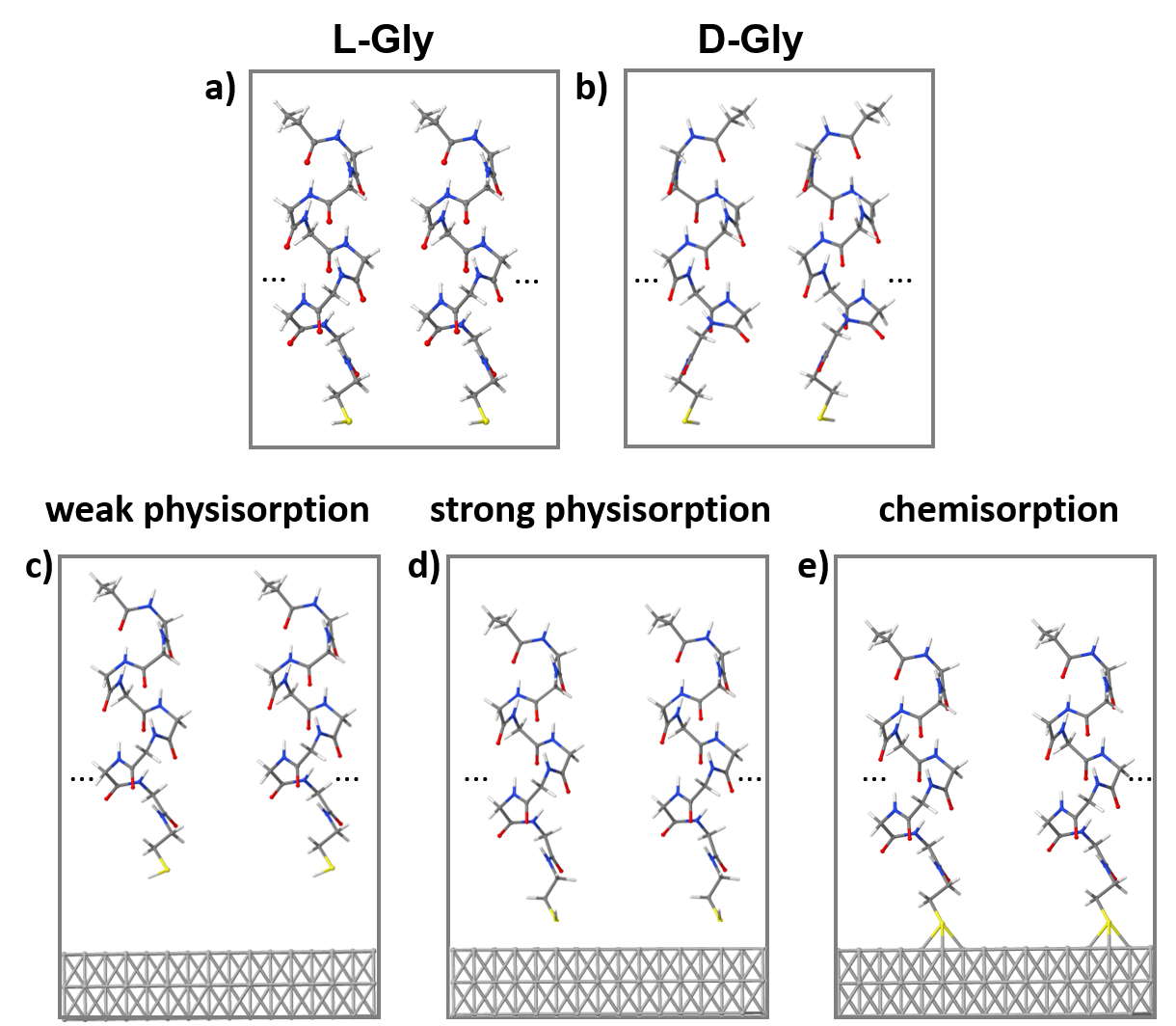}
    \caption{Panels a) and b) show the investigated molecular systems: L- and D-glycine peptides with a length of approximately 15 \AA\,. By exploiting periodic boundary conditions, a two-dimensional molecular array with varying packing density can be mimicked. Panels c)-e) show  three different regimes of interaction (labeled by the {vertical} distance $d_{S-Ni}$) between the  sulfur atom belonging to  the thiol linker of the oligopeptide and the Ni(111) surface: (c) weak physisorption ($d_{S-Ni}$=5 {\AA}), (d) strong physisorption ($d_{S-Ni}$=1.8 {\AA}), (e)  and chemisorption ($d_{S-Ni}$=1.6 {\AA}). }
    \label{fig:1}
\end{figure}

We have chosen as a reference molecular system a polyglycine $\alpha$-helix with its D-(D-Gly) and L-(L-Gly) enantiomers and a length of  15 \AA. This oligomer length  is roughly of the same order of magnitude as in some experimental studies,~\cite{doi:10.1021/acs.jpclett.9b01433} but shorter as in AFM experiments.~\cite{doi:10.1002/adma.201904206} The two top panels of Figure~\ref{fig:1} show the two enantiomers. Since we are going to also study interactions with a substrate, one of the termini of the molecules was functionalized with a thiol SH-group.

We mention that it is known that polyglycine does not build stable helical structures at ambient conditions. However, since we are interested in very generic features related to helical molecules, we consider it as an atomistic toy model to explore the role of exchange effects. Moreover,  the main difference to the experimentally studied polyalanine oligomers is the presence, in the latter, of methyl groups, which are not expected to play any fundamental role in the problem we are dealing with, since they should not display any sizeable spin density. 

The first part of this study is motivated by   the enantiomer separation studies mentioned in the introduction.~\cite{Banerjee-Ghosh1331}  Due to the complexity of the problem, we are going to consider different idealized cases, which will,  nevertheless, allow us to \change{distill}  the influence of exchange interactions. We  consider the following situations: (i) periodic two-dimensional molecular arrays without a substrate,  consisting only of  D-Gly or L-Gly. The first question to address is, how far  intermolecular interactions can modify the electronic structure of the array, and which  is the critical separation (if any), i.e. molecular packing density, between neighboring molecules leading to a qualitative change of it.  Once this critical inter-molecular distance has been spotted, the interaction of the array with a magnetized Nickel substrate will be investigated for three different regimes: weak physisorption (large assembly-surface separation), strong physisorption (intermediate assembly-surface separation) and chemisorption, involving a chemical bond of the thiol linker to a Ni atom on the surface. These three situations are displayed in the lower panel of Figure~\ref{fig:1} for the case of D-Gly. We proceed now to discuss these individual situations in more detail. 

%\paragraph*{(i) Isolated molecular arrays}
{\textit{(i) Isolated molecular arrays.}}  We start our investigation with  molecular assemblies in the absence of any substrate. For this, we use a simulation cell containing a single molecule  with periodic boundary conditions in x- and y-directions (perpendicular to the molecular axis) and a vacuum gap of 15 \AA\, in z-direction to avoid spurious interactions. In order to mimic different packing densities, we have systematically varied the edge lengths  of the simulation cell in x- and y-direction (denoted by $L_{x}=L_{y}=L_0$) in the range from 20 {\AA} to 8 {\AA}, the latter case roughly corresponding to a molecular number density slightly smaller than $1\times 10^{14}$ mol/cm$^{2}$. This order of magnitude agrees well with experimental estimates.\cite{doi:10.1021/jp050398r,Ray99,doi:10.1021/acs.jpcb.9b07987}

\begin{figure}[t!]
    \centering
    \includegraphics[width=0.98\textwidth]{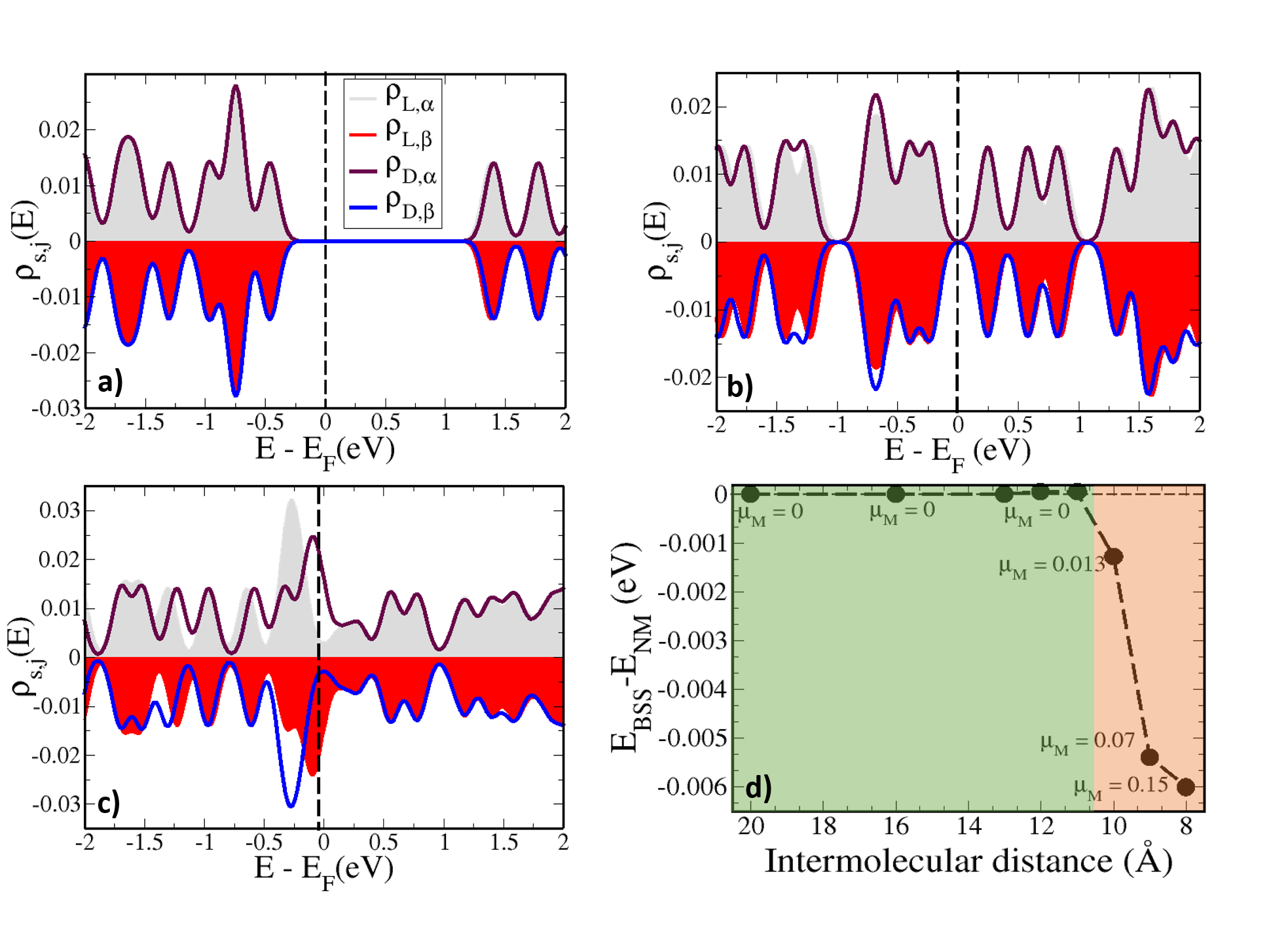}
    \caption{Panels a)-c):  Total spin-resolved density of states $\rho_{s,j}$, 
    with $s$=D,L and spin component $j=\alpha,\beta$, of helical polyGly for 
    different  inter-molecular separations determined by the edge length $L_0$ of
    the simulation box. a) $L_0$=20{\AA}, b) $L_0$=13{\AA}, and c) $L_0$=8{\AA}. For case c) an insulator-to-metal 
    transition takes place, accompanied by the emergence of a  spin asymmetry of both enantiomers, mostly visible around the Fermi energy. Since each 
    enantiomer has two possible energetically degenarate Broken Symmetry 
    Solutions (BSS), we only show one of them for D- and L-Gly, respectively. Panel d) shows the energetic difference 
    between the BSS and the non-magnetic (NM) solution for  L-Gly as a 
    function of the intermolecular separation. For the curve shown, the initial 
    magnetic moment per atom was positive. The non-magnetic state (with non-zero 
    band gap) remains energetically degenerate with the BSS state down to a 
    distance $\le 10 \AA$; for smaller separations the system metallizes and, 
    additionally, the BSS becomes energetically favored. By choosing the initial 
    magnetic moment per atom negative, a similar behavior is obtained, but now 
    with $\mu_{M} < 0$ (not shown). The same result is found  for the 
    D-enantiomer.}
    \label{fig:2}
\end{figure}

\change{In order to explore the possible magnetic solutions, we have used for each $L_0$ and for each enantiomer  (L or D) two different initial conditions, denoted by $m_{up}$ and $m_{dw}$, in the spin-resolved DFT calculations. These variables  initialize  the magnetic moment per atom in the system 
%(technically, this corresponds to two possible choices of the variable MAGMOM in the VASP code). 
The system is then allowed to relax to find the energetically more favorable electronic configuration. As far as the distance between the molecules in the array is larger than a critical value (in our case $\sim 10$\AA) the magnetic ground state for both enantiomers is a singlet with no broken spin symmetry. This is shown in Figure~\ref{fig:2}a, where the total spin density of states (DOS) $\rho_{s,j}$ on the molecule is  shown for the case $L_0$=20 \AA. The index $s$=D,L labels the enantiomer type and $j=\alpha,\beta$, labels the spin components. The electronic band gap in this case is approximately 1.4~eV for both helicities. This case basically corresponds to isolated molecules with $\rho_{s,\alpha}=\rho_{s,\beta}$. This result is expected, since isolated molecules should not display any asymmetries in their spin densities. This is further illustrated by looking at the spin-resolved DOS projected on selected atoms of the helix (shown in Figure~SI2 in the Supplementary Section), which also shows perfect symmetry for both spin components.  When the parameter $L_0$ becomes smaller, see Figure~\ref{fig:2}b for the case $L_0$=13 \AA, the band gap starts to get narrower (few meV), although no significant asymmetries in the spin densities for both enantiomers are yet visible, implying that the ground state has still no broken spin symmetry. 
 
However, when $L_0\sim 10$ \AA\, or smaller, a dramatic change in the electronic structure takes place, as seen in Figure~\ref{fig:2}c. In this case, the band gap completely closes, leading to an insulator-to-metal like electronic phase transition. Additionally,  $\rho_{s,\alpha}\neq\rho_{s,\beta}$, mostly in an energy window around the Fermi energy. This indicates a broken spin symmetry state (BSS), despite the fact that we still have a singlet ground state.~\cite{doi:10.1021/jp054583g,doi:10.1021/jz201326k} In fact, if each component of the DOS is integrated over the full energy range, the same value of the total spin DOS corresponding to a singlet S=0 (within numerical accuracy) is obtained. We  also stress that there is no breaking of time-reversal symmetry in this situation, since there are no external  magnetic fields or magnetic exchange interactions present.  

An important point is that,  for a given enantiomer, two possible magnetic solutions are found in dependence on the chosen initial conditions   $m_{up}$ or  $m_{dw}$, and giving positive (plus) and negative (minus)  net spin densities around the Fermi energy, respectively. Both solutions are energetically degenerate as it should hold for a system not influenced by  spin-sensitive perturbations. These solutions correspond to the BSS states previously mentioned for the case of gold nanoparticles. Notice that in Figure~\ref{fig:2}c we have shown the specific case of the minus-solution for  L-Gly and the plus-solution for  D-Gly. In Figure SI1 in the Supplementary Section we show the full set of solutions depending on the initial magnetic moment per atom. As shown in Figure SI3 in the Supplementary Section, the largest contribution to the spin DOS  around the Fermi energy  arises from contributions of  oxygen atoms and from the sulfur atom in the thiol group. Associated with the observed spin asymmetry there is also a magnetic moment of 0.15 (-0.15) $\mu_B$ for the plus and minus solutions, respectively.  

Figure~\ref{fig:2}d summarizes the previously discussed behavior. We plot there the energy difference between the broken symmetry solution and the non-magnetic  solution, $E_{BSS}-E_{NM}$, as a function of the intermolecular distance in the 2D array. It becomes clear from this figure, that once the critical separation is reached, the broken symmetry state becomes energetically favored and a non-zero magnetic moment $\mu$ emerges. As previously mentioned, there is another solution with $-\mu$ and with the same energy (not shown in the Figure). This scenario is valid for both enantiomers. 

Interestingly, the described transition is not present in the case of linear peptides. As shown in Figure SI4 in the Supplementary Section, in a 2D array of linear glycine peptides not even a metallization of the array can be found for intermolecular separations down to $\sim 5$ \AA\ and the nonmagnetic singlet state is always the ground state solution. In related systems, the closing of the gap as a result of the  electronic phase transition is controlled by the relative magnitude of exchange effects and dipole moments.~\cite{C8CP04208J} However, both the helical and the linear molecules have non-zero dipole moments. This hints to a strong dependence on the range of intermolecular interactions, which are clearly enhanced in the helical case, where molecules can interpenetrate each other, compared to the linear one, where interpenetration is hampered. In this sense, the behavior found in the helical peptide seems to arise from an interplay of helical shape and exchange effects. The closing of the band gap below the critical distance is clearly seen in the band structure plot in the same figure. The effect is also less sensitive to the specific linker. Thus, thiol groups and carboxylic groups yield qualitatively similar results (see Figure SI5 in the Supplementary Section).

As previously mentioned,  the main contributions to the computed  molecular magnetic moment in the closed-packed array   largely arise from the sulfur atom and neighboring oxygens atoms. This can  provide a basis to justify the formulation of  Heisenberg-type model Hamiltonians describing the exchange interaction between a nearly localized spin density on the helical molecule and a magnetized substrate. These models may allow to address, in future studies, in a more specific, though phenomenological way the kinetics of the interaction between helical molecules and magnetized surfaces.
}

%\paragraph*{(ii) Exchange interaction with surfaces}
\textit{(ii) Interaction with surfaces.}
\change{The previous discussion hints at a non-trivial magnetic response of helical molecular arrays. Isolated arrays build obviously an idealization and no difference between L- and D-enantiomers can be seen. To explore the possible influence of a substrate, } we include now  a magnetized Ni(111) surface to address the  interaction with the molecular array. Since from our previous analysis the most interesting case is the one shown in Figure~\ref{fig:2}c for $L_0$=8 \AA, we will only consider this case in the following (due to the inclusion of a substrate and the used periodic boundary conditions, the actual intermolecular separation in the supercell is slightly larger than 8 \AA). 

We have   treated three  adsorption stages, according to the {vertical} distance between the nickel surface and the sulfur atom in the thiol group, as shown schematically in Figure~\ref{fig:1}c-e:  weak physisorption (d$_{S-Ni}$=5 \AA), strong physisorption (d$_{S-Ni}$= 1.8 \AA), and chemisorption (d$_{S-Ni}$=1.6 \AA). In this latter case, d$_{S-Ni}$ is smaller than the typical values obtained in experimental studies~\cite{doi:10.1021/acs.langmuir.5b00177}, since they would need to be corrected to account for the tilting of the molecules once they are linked to the substrate.  Note also that in the cases of weak and strong physisorption, the hydrogen atom is still bonded to the sulfur atom, whereas it is removed for the chemisorbed case.  The three bonds seen in the chemisorption case result from the sulfur atom being positioned on a hollow site. The magnetization direction of the Ni substrate was perpendicular (out of plane) to the surface direction in all cases discussed in this section. For these calculations, we have used a 4$\times$4 supercell (with dimensions 9.96 \AA $\times$ 8.62 \AA $\times$ 43 \AA). We only consider here the case of the substrate magnetized with  magnetic moments parallel to the z-direction, the   anti-parallel case is shown in Figure \change{SI6} of the Supplementary Section, showing just a sign inversion of the spin densities.

\begin{figure}[t!]
    \centering
    \includegraphics[width=0.99\textwidth]{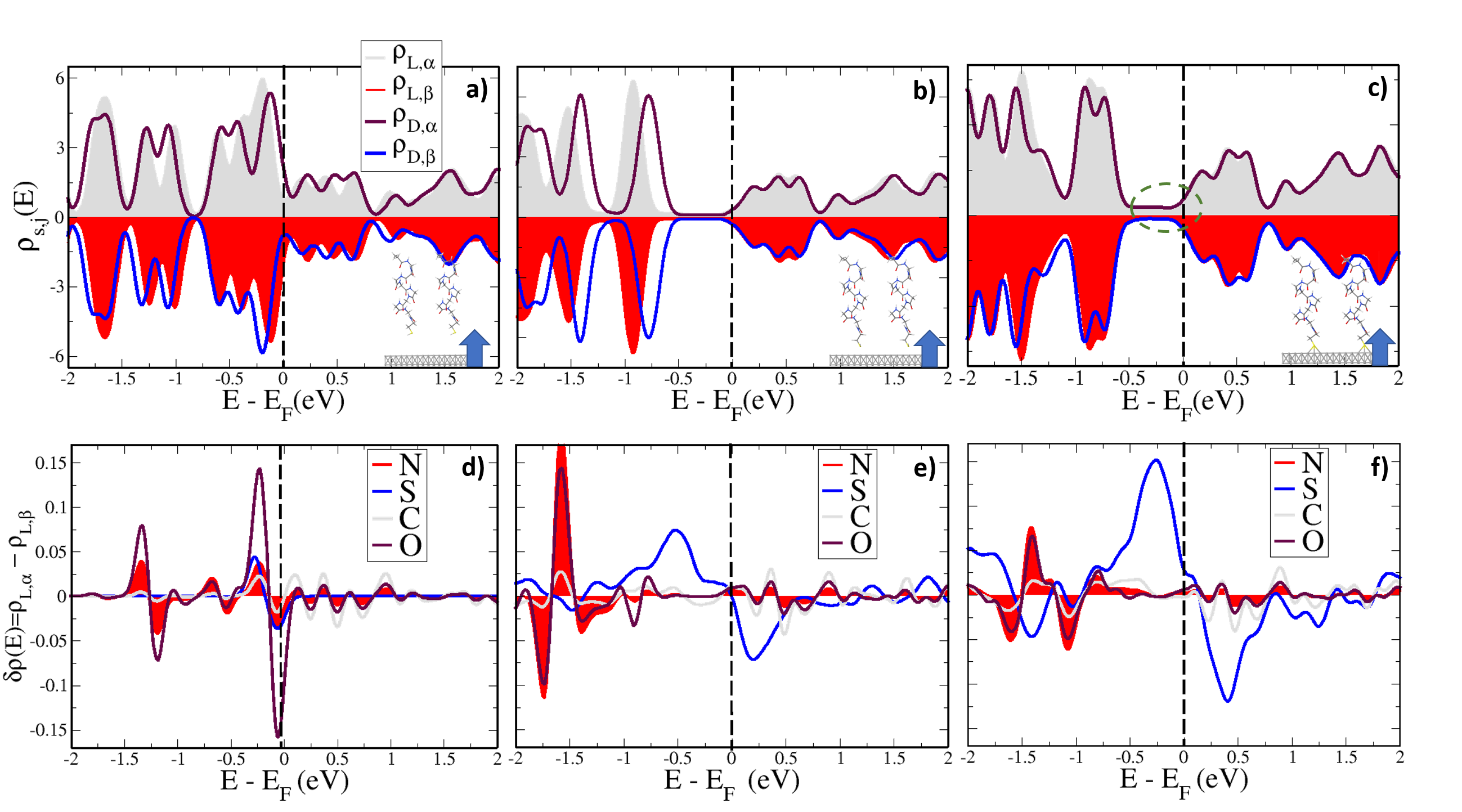}
    \caption{Total spin-resolved density of states $\rho_{s,j}$, with $s$=D,L and spin component $j=\alpha,\beta$  for different molecule-surface distances with a given magnetization direction of the substrate (vertical arrow in the insets). The three cases considered are a) weak physisorption, b) strong physisorption, and c) chemisorption. With decreasing molecule-surface distance the band gap reopens, although it remains small. Any sizeable difference between  L- and D-enantiomers  disappears for the chemisorption case. A small non-zero spin density remains, however,  around the Fermi energy in this latter case (enclosed region in panel c). Panels d)-f) show the net projected spin density of states ($\delta\rho(E)=\rho_{L,\alpha}-\rho_{L,\beta}$) on different atoms in L-Gly for the three cases depicted in panels a)-c). While in the weak physisorption case (panel d) the dominant contribution to the spin density around the Fermi energy arises from the oxygen atoms, the largest contribution in the chemisorption case (panel f) is now   coming from the sulfur atom which binds to the surface.
    }
    \label{fig:3}
\end{figure}

Figure~\ref{fig:3}a-c shows the spin-resolved density of states $\rho_{s,j}$ for the three cases. As expected, for weak physisorption, the same results are obtained as for the case of  an array in the absence of a surface. \change{Notice that we have now to present the solutions for L-Gly and D-Gly for the same initial choice of magnetic moment, i.e. $m_{up}$, since this is the selected state of the Ni surface}.  For strong physisorption (Figure~\ref{fig:3}b), on the contrary, the gap reopens and any spin asymmetries is now considerably reduced for both enantiomers. 

Finally, once the molecule is chemisorbed, the difference between enantiomers is largely washed out ($\rho_{D,j}\approx\rho_{L,j}$); however, there is still a net spin density  around the Fermi energy (enclosed region in Figure~\ref{fig:3}c) with spin-up dominating for both enantiomers. The DOS in the gap region results from hybridization between the sulfur and nickel electronic states.  

In Figure~\ref{fig:3}d-e the atomic-resolved net spin density $\delta\rho=\rho_{L,\alpha}-\rho_{L,\beta}$ for the L-Gly molecule is shown for N, S, C, and O atoms. For the weak physisorption regime, contributions from the oxygen atoms and, to a lesser degree, from S,C, and N, are dominant around the Fermi energy. The situation changes, however, for the chemisorption case, where the sulfur atom provides the largest contribution to the spin density  around the Fermi energy, while all the other contributions (O,C,N) are largely depleted.

\change{Since we do not have a physical magnetic field in our treatment, no real breaking of symmetry between the L- and D-enantiomers can be obtained, i.e. states with positive and negative net spin density around the Fermi energy for each enantiomer remain degenerate. To highlight the potential effect of an external magnetic perturbation, we have considered a simplistic situation, where a single Ni atom has been placed close to the thiol terminus of the molecules in the weak physisorption case.  The results, displayed in Figure SI7 of the Supplementary Section, clearly show a symmetry breaking, the net spin density becoming now larger and positive for  D-Gly when the intial magnetic
moment of the system is up, while it is larger and negative for L-Gly when the intial magnetic moment is down. 

Although a direct comparison with experimental results is obviously not straightforward, the discussed behavior  with decreasing separation from the substrate as well as the previously described symmetry breaking between L-Gly and D-Gly in presence of an external perturbation (mimicking a magnetic field)  provides a possible  way to rationalize   the experimental results in Ref.~\cite{Banerjee-Ghosh1331}, where enantiomeric selectivity is lost   once the molecules bind to the substrate and the systems enter into the equilibrium thermodynamic regime. 
}

\begin{figure}[t!]
    \centering
    \includegraphics[width=0.98\textwidth]{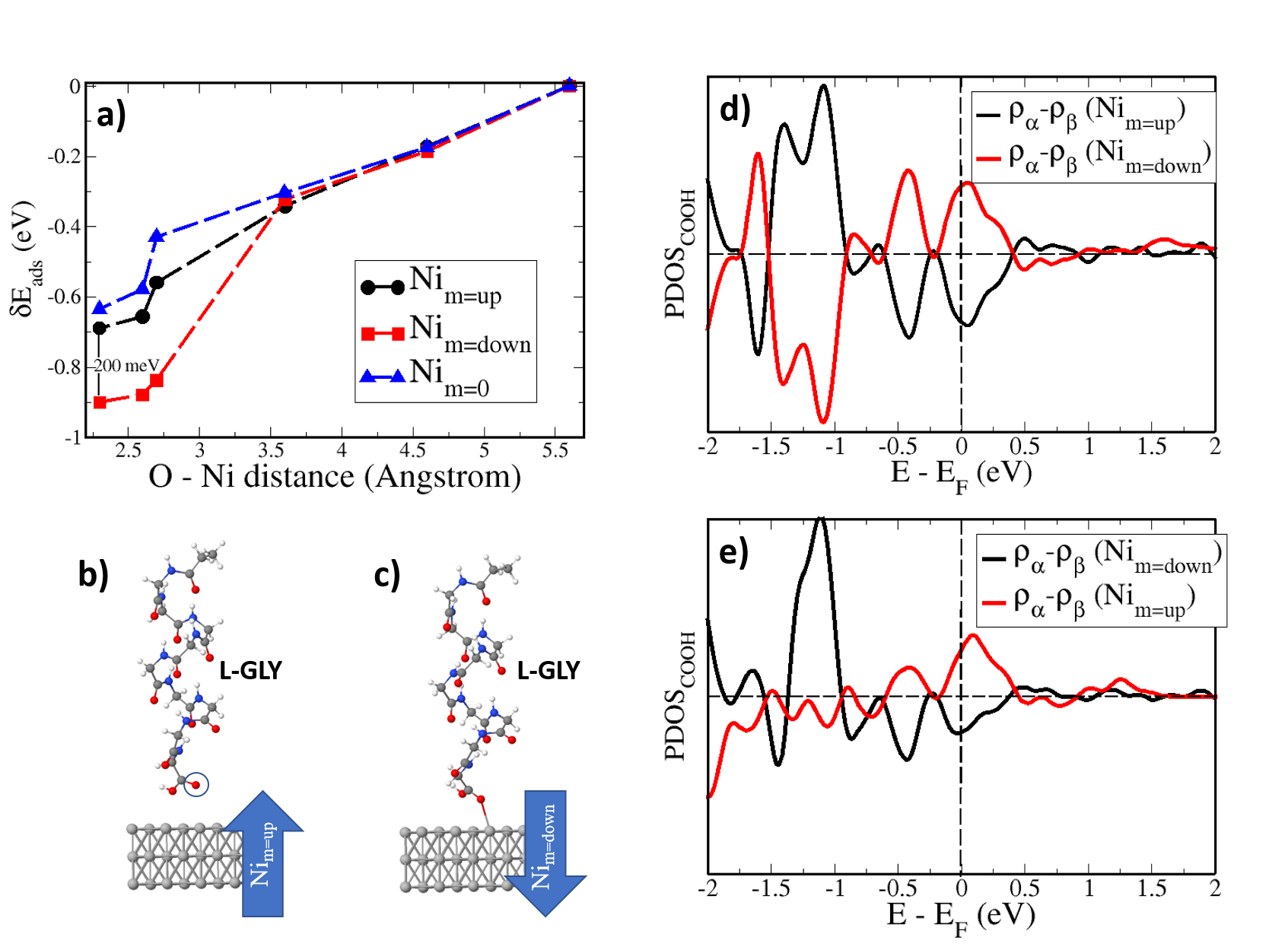}
    \caption{a)  Adsorption energy $\delta E_{\textrm{ads}}$ of the $\alpha$-helix as a function of the vertical distance to the surface for different surface magnetization directions as well as for a non-magnetized substrate. 
    As (arbitrary) reference energy the one for an O-Ni distance of 5.60 \AA\, was chosen, i.e. $\delta E_{\textrm{ads}}=E_{d}-E_{d=5.60}$. Stronger adsorption (largest energy change) was found for the case of down surface magnetization, while for a non-magnetized surface, the adsorption energy is weakest. For the shortest studied O-Ni separation, an exchange energy of the order of 200 meV can be estimated.
    b) Optimized geometry of the system for the up magnetization of the substrate for a vertical separation of 2.6 \AA between the (encircled) oxygen atom and the surface. c) Optimized geometry for the same O-Ni separation, but with down magnetization direction of the substrate. A much stronger structural relaxation takes now place in the molecule. 
    Panels d) and e) show the net spin density $\rho_{ \alpha}-\rho_{ \beta}$), projected on the carboxylic group for both magnetization orientations of the substrate. In d) a single point calculation using the same molecular structure for up- and down-magnetization directions was carried out and as a result, a  totally symmetric net spin density upon reversing the substrate magnetization is obtained. In e) structural relaxation upon inversion of the magnetization direction of the surface is allowed, leading now to stronger asymmetries.}
    \label{fig:4}
\end{figure}

\textit{Exchange interactions with a magnetized substrate: mimicking magnetic exchange AFM setups.} In this last section, we address a situation similar to that of Ref.~\cite{doi:10.1002/adma.201904206}. The simulation setup is similar to that previously used when discussing the results of Figure~\ref{fig:3}. The only difference now is that we use here, similar to Ref.~\cite{doi:10.1002/adma.201904206}, a carboxylic COOH-group at the termini facing the substrate, instead of a thiol linker. We remark that using a COOH group also leads to an insulator-to-metal transition in closely packed arrays as it happened for the case of thiol groups (see Figure SI5 in the Supplementary Section). 
We further consider the phsysorption regime and two possible magnetization directions of the Ni(111) substrate, parallel or anti-parallel to the z axis. We only discuss the L-GLY molecule and compute the total energy of the system as a function of the perpendicular distance between an oxygen atom belonging to the COOH-group (encircled atom in Figure~\ref{fig:4}b) for both magnetization directions of the substrate as well as for the case of a non-magnetized surface. Structural relaxation is performed for each given  separation, which ranges from  5.6 \AA\, down to 2.3 \AA. In Figure~\ref{fig:4}a, we show the corresponding energy-vs-distance curves, taking as a (arbitrary) reference energy the one for a vertical O-Ni distance of 5.60 \AA, i.e. $\delta E_{\textrm{ads}}=E_{d}-E_{d=5.60}$. It becomes clear from the figure that upon approaching the substrate down to a distance of roughly 3.60 \AA, no significant  differences in the total energies of the system for the three magnetic states of the substrate exists. However, below this distance and down to 2.3 \AA\,  (reaching  the strong physisorption regime), an increasing energy difference between the up- and down-magnetic configurations develops, with the down configuration being energetically favored. The main reason for this behavior is related to structural relaxation effects, as illustrated in Figure~\ref{fig:4}b and Figure~\ref{fig:4}c. While for the up-magnetization direction of the substrate only very weakly structural changes of the molecule are found, stronger structural relaxation takes place for the down-magnetization direction, with one oxygen of the carboxylic group coming closer to the surface. 
This results in a change in the vertical distance of the oxygen atom (labeled in Figure~\ref{fig:4}b), which gets by about 0.8 \AA\, closer to the surface for the down magnetization direction as compared to the up magnetization direction. For the shortest separation of 2.3 \AA the difference in energy between both orientations amounts  200 meV, which provides a measure of the exchange interaction between the molecule and the substrate. Taking into account the strong simplifications of our modelling setup, this value agrees  qualitatively well with the estimation of Ref.~\cite{doi:10.1002/adma.201904206} based on the experimental data:  by calculating   the breaking energy and integrating  the  measured force-distance curve  from  the  pulling  position  to  the  zero-force  position sensed by the cantilever, the authors obtained an estimate of 150 meV. To confirm that (exchange-mediated) structural relaxation is an important  factor determining the obtained energy difference, we show in panels d) and e) of  Figure~\ref{fig:4} 
the net spin density $\rho_{\alpha}-\rho_{\beta}$, 
projected on the carboxylic group for both magnetization orientations of the substrate. Figure~\ref{fig:4}d is a single point calculation using the same molecular geometry for {\textit{both}} magnetization directions, i.e. without allowing for structural relaxation upon switching of the substrate magnetization direction. Figure~\ref{fig:4}e shows, on the other hand,  the case with structural relaxation. It is evident, by comparing both panels, that the  spin density is strongly affected by structural relaxation and that it displays a largely different behavior around the Fermi energy when switching the magnetization direction.  This result also agrees well with the  experimental studies, where a larger pulling force was found for a down substrate magnetization (see Figure 2a in~\cite{doi:10.1002/adma.201904206}).

%\textcolor{red}
{To support the theoretical results above, a new analysis of the force-distance curves from Ref.~\cite{doi:10.1002/adma.201904206} was performed  using Principal Component Analysis (PCA),~\cite{pca} see the Supplementary Section for additional details.
By inspecting the first two principal components (PC1 and PC2), which span 50\% of the data variance, new insights can be drawn. In Figure~SI9 of the Supplementary Section it is demonstrated that both up and down magnetizations differ from the nonmagnetic substrate in the PC space and that the difference is due to a significant contribution of PC1 (stronger adhesion) and a reduced  contribution of PC2
(weak adhesion). This new analysis points to two different adhesion mechanisms, one related to spin exchange (PC1) and one that does not involve such spin dependence (PC2). These results correlate very well with the theoretical calculations presented in Figure~\ref{fig:4}, where the surface magnetization increases the binding energy between the molecule and the substrate due to structural relaxation, a mechanism relevant only for the magnetized surface.}

Calculations for the chemisorbed case are shown in Figure~SI8 of the Supplementary Section. We find that the local binding structure at the molecule-surface interface is very similar for both magnetization directions, leading to a total energy difference of only $\sim$ 20 meV between up- and down magnetization directions. On the contrary, the energy of the system for a non-magnetized substrate is considerably larger and amounts to 180 meV, indicating that the preferred states are those with magnetized substrates.

%\textcolor{red}
{These results may provide a plausible explanation for the lack of contrast between up and down magnetization observed in the experimental studies (see in particular Figure 4c in Ref.~\cite{doi:10.1002/adma.201904206})  upon keeping the AFM tip on the surface for 1 sec before retracting. Through this process, the probability of covalent bond formation is increased, thus resulting in a reduced contrast between the up and down magnetization directions as our  calculations above illustrate.}

In conclusion, we have addressed, on a first-principle basis, the question of the potential role of exchange interactions in influencing various spin-dependent effects in helical molecules as recently observed in experimental investigations.~\cite{Banerjee-Ghosh1331,doi:10.1063/1.5125034,doi:10.1002/adma.201904206} \change{Our main finding is the possibility of obtaining a broken spin symmetry (singlet) state resulting from inter-molecular  interactions in arrays of helical molecules. The fact that arrays of linear peptides did not show any spin asymmetries (and not even a transition to a  metallic state) at short inter-molecular separations is a strong hint at the non-trivial role of the helical shape for the magnetic response of these molecules. Clearly, the issue of the influence of a physical strong magnetic field or short-range spin-spin exchange interactions remains open and needs to be considered in order to quantitatively address the  breaking of the symmetry between L- and D-enantiomers. }
We finally  remark that   preliminary calculations including spin-orbit coupling (not included in this manuscript) did not show any sizeable effects in the main features  discussed in the current investigation. This may be an indication that spin-orbit effects gain in significance when considering charge transport or transfer through the helical molecules, but eventually  play a less essential role in the type of setups investigated here. 
\change{This also points at the need to reconsider the standard expressions for van der Waals forces involving chiral molecules, because in addition to the conventional description that considers only electric dipole mediated  contributions, it would be necessary to consider the induced magnetic dipoles related to the exchange interaction. This is an exciting topic on which we are currently working}.

 \begin{suppinfo}
%
%  A listing of the contents of each file supplied as Supporting Information
%  should be included. For instructions on what should be included in the
%  Supporting Information as well as how to prepare this material for
%  publications, refer to the journal's Instructions for Authors.
%
The following files are available free of charge.
\begin{itemize}
    \item  Dianat-etal-SupplementarySection.pdf.
\end{itemize}
\end{suppinfo}
%%%%%%%%%%%%%%%%%%%%%%%%%%%%%%%%%%%%%%%%%%%%%%%%%%%%%%%%%%%%%%%%%%%%%
%% The "Acknowledgement" section can be given in all manuscript
%% classes. This should be given within the "acknowledgement"
%% environment, which will make the correct section or running title.
%%%%%%%%%%%%%%%%%%%%%%%%%%%%%%%%%%%%%%%%%%%%%%%%%%%%%%%%%%%%%%%%%%%%%
\begin{acknowledgement}
The authors thank Elena Diaz and Francisco Dominguez-Adame for very fruitful discussions. V. M. thanks Eduardo V. Lude\text{\~{n}}a for very fruitful discussions concerning exchange effects. 
A. D. and G. C. acknowledge financial support from the Volkswagen Stiftung (grant nos. 88366). This project has received funding from the
    European Union’s Horizon 2020 research and
    innovation programme under the Marie
    Skłodowska-Curie grant agreement No 813036. We
acknowledge the Center for Information Services and High
Performance Computing (ZIH) at TU Dresden for providing the necessary computational resources. Vladimiro Mujica acknowledges a Fellowship from Ikerbasque, the Basque Foundation for Science.
\end{acknowledgement}

%%%%%%%%%%%%%%%%%%%%%%%%%%%%%%%%%%%%%%%%%%%%%%%%%%%%%%%%%%%%%%%%%%%%%
%% The same is true for Supporting Information, which should use the
%% suppinfo environment.
%%%%%%%%%%%%%%%%%%%%%%%%%%%%%%%%%%%%%%%%%%%%%%%%%%%%%%%%%%%%%%%%%%%%%

%\bibliography{paper}

\providecommand{\latin}[1]{#1}
\makeatletter
\providecommand{\doi}
  {\begingroup\let\do\@makeother\dospecials
  \catcode`\{=1 \catcode`\}=2 \doi@aux}
\providecommand{\doi@aux}[1]{\endgroup\texttt{#1}}
\makeatother
\providecommand*\mcitethebibliography{\thebibliography}
\csname @ifundefined\endcsname{endmcitethebibliography}
  {\let\endmcitethebibliography\endthebibliography}{}

%\end{multicols}{2}

%%%%%%%%%%%%%%%%%%%%%%%%%%%%%%%
%%%%%%%%%%%%%%%%%%%%%%%%%%%%%%%

%%%%%%%%%%%%%%%%%%%%%%%%%%%%%%%
%%%%%%%%%%%%%%%%%%%%%%%%%%%%%%%%%%%%%

%%%%%%%%%%%%%%%%%%%%%%%%%%%%%%%%%%%%%
%%%%%%%%%%%%%%%%%%%%%%%%%%%%%%%%%%%%%%%%

\end{document}